\documentclass{raa}

\usepackage{graphicx,times}             %for PS/EPS graphics inclusion, new
\usepackage{amssymb}
\input{epsf.sty}                        %for PS/EPS graphics inclusion, old

\begin{document}

   \title{The LEGUE Input Catalogue for Dark Night Observing in the LAMOST Pilot Survey}

   \volnopage{Vol.0 (200x) No.0, 000--000}      %%preserved for Editor. DOn't remove!
   \setcounter{page}{1}          %%starting page, preserved for Editor. DOn't remove!

   \author{ Fan Yang \inst{1,3} \and Jeffrey L. Carlin \inst{2} \and Chao Liu \inst{1} \and  Yueyang Zhang \inst{1,3} \and Shuang Gao \inst{1} \and Yan Xu \inst{1} \and Licai Deng \inst{1} \and Heidi Jo Newberg \inst{2} \and S\'ebastien L\'epine \inst{4} \and Jinliang Hou \inst{5} \and Xiaowei Liu \inst{6} \and Norbert Christlieb \inst{7} \and Haotong Zhang \inst{1} \and Hsutai Lee \inst{8} \and Kaike Pan \inst{9} \and Zhanwen Han \inst{10} \and Hongchi Wang \inst{11}}

   \institute{     Key Lab for Optical Astronomy, National Astronomical Observatories, Chinese Academy of Sciences, Beijing 100012, China ({\it fyang@LAMOST.org})\\
   	 \and        Department of Physics, Applied Physics, and Astronomy, Rensselaer Polytechnic Institute, 110 8th Street, Troy, NY 12180, USA\\
        \and        Graduate University of Chinese Academy of Sciences, Beijing 100049, China\\
        \and        American Museum of Natural History, Central Park West at 79th Street, New York, NY 10024, USA\\
        \and        Shanghai Astronomical Observatory, Chinese Academy of Sciences, 80 Nandan Road, Shanghai 200030, China\\
        \and        Department of Astronomy \& Kavli Institute of Astronomy and Astrophysics, Peking University, Beijing 100871, China\\
        \and        University of Heidelberg, Landessternwarte, Kšnigstuhl 12, D-69117 Heidelberg, Germany\\
        \and        Academia Sinica Institute of Astronomy and Astrophysics, Taipei, China \\
        \and        Apache Point Observatory, 2001 Apache Point Road
P.O. Box 59, Sunspot, NM, USA\\
        \and        Yunnan Astronomical Observatory, Chinese Academy of Sciences, Kunming 650011, China\\
        \and        Purple Mountain Observatory, Chinese Academy of Sciences, Nanjing, Jiangsu 210008, China\\}
   \date{Received~~2012 month day; accepted~~2012~~month day}

\abstract{ We outline the design of the dark nights portion of the LAMOST Pilot Survey, which began observations in October 2011. In particular, we focus on Milky Way stellar candidates that are targeted for the LEGUE (LAMOST Experiment for Galactic Understanding and Exploration) survey. We discuss the regions of sky in which spectroscopic candidates were selected, and the motivations for selecting each of these sky areas. Some limitations due to the unique design of the telescope are discussed, including the requirement that a bright ($V < 8$) star be placed at the center of each plate for wavefront sensing and active optics corrections. The target selection categories and scientific goals motivating them are briefly discussed, followed by a detailed overview of how these selection functions were realized. We illustrate the difference between the overall input catalog -- Sloan Digital Sky Survey (SDSS) photometry -- and the final targets selected for LAMOST observation.
}

   \authorrunning{F. Yang, J.~L. Carlin, \& L. Chao, et al.}            %author_head in even pages
   \titlerunning{LAMOST/LEGUE Dark Nights Observing Program }  % title_head in odd pages

   \maketitle
   
\section{Introduction}           
\label{sect:intro}  

The LAMOST survey, slated to begin in late 2012 (see Zhao et al. 2012 for an overview), will obtain spectra of millions of Milky Way stars (in a survey known as LEGUE -- LAMOST Experiment for Galactic Understanding and Exploration) in addition to an extragalactic survey of QSOs, galaxy redshifts, and stellar populations of galaxies (known as LEGAS -- LAMOST ExtraGAlactic Surveys). 
%The unique design of the 4-meter effective aperture LAMOST telescope and its instruments provides an unprecedented opportunity for studies of the structure of our Galaxy. 
The LAMOST telescope has a $5^\circ$-diameter field of view, and a focal plane populated with 4000 robotically-positioned optical fibers feeding 16 bench spectrographs. The effective aperture of the telescope varies from 3.6 to 4.9 meters depending on the pointing position (Cui et al. 2012). The combination of a large field of view, ample collecting area, and highly-multiplexed spectroscopy enables surveys of orders of magnitude more targets than was previously feasible, and covering huge contiguous areas of sky. Thus LAMOST (in particular, the LEGUE Galactic structure portion of the survey) will provide unprecedented opportunities for studies of the structure of our Galaxy. 

With these new capabilities (and limitations imposed by the unique telescope design that enabled the large-scale spectroscopic survey) come new challenges in selecting input catalogs for observing. We report here on the design of the LAMOST Pilot Survey, a one-year program operated in survey mode, which will provide a rich spectroscopic data set that allows testing of survey mode operations (in addition to producing science results) before the main LAMOST survey begins. The LAMOST Pilot Survey began on 24 October 2011, and will operate through the end of spring 2012. 

In order to more efficiently take advantage of telescope time during all observing conditions, two survey modes were adopted for the Pilot Survey. Separate input target catalogs were generated for observations on bright and dark nights. Bright nights are defined as from 5 nights before to 5 nights after the full moon in each lunation.  Dark nights are those from 5 nights before to 5 nights after the new moon, and those between the dark and bright are defined as grey time. During the Pilot Survey, six nights are set aside for test observations and telescope adjustments per lunar month from 5-7 nights and from 20-22 nights after the new moon. On the bright nights, relatively bright ($r \lesssim 16.5$) stars are observed in the low-latitude regions near the Galactic anti-center (GAC), the Galactic disk area, and a constant-declination stripe at $\delta \sim 29^\circ$. Faint targets are reserved for dark nights, when the LEGUE survey of the Galactic halo and the LEGAS extragalactic surveys are carried out, and portions of the grey nights when moonlight is minimal. In this paper, we discuss the design of the dark nights portion of the LAMOST Pilot Survey; a companion paper (Zhang et al. 2012) will discuss the bright nights observing program.
%part (known as LEGUE, LAMOST Experiment for Galactic Understanding and Exploration, not including GAC?) and extragalactic objects (known as LEGAS, LAMOST ExtraGAlactic Surveys) are carried. The pilot survey will tell us how this telescope can perform and what kind of scientific goal we can achieve during the main survey.

In this paper we discuss the design of the stellar input catalogue for dark nights, including the areas of sky to be observed and special considerations required due to site and telescope limitations and the sharing of plates with the LEGAS survey. Because LEGAS cannot fill all of the fibers on each plate with extragalactic objects, many fibers on LEGAS plate are available for LEGUE targets. We discuss some details of the selection of individual targets for the dark nights portion of the LAMOST Pilot Survey; details of target selection algorithm and the bright nights and disk surveys are given elsewhere (Carlin et al. 2012, Zhang et al. 2012, Chen et al. 2012). This paper is organized as follows:  In Section 2, we introduce how the observation area for dark nights was selected. Section 3 describes how we selected the targets. Section 4 describes the plate design. We conclude with some brief discussion in Section 5.

%LAMOST is a quasi-meridian reflecting Schmidt telescope laid down on the ground with its optical axis fixed in the meridian plane, which has 4000 fibers on the focal plane and has a $5^\circ$ field of view. The pilot survey of LAMOST started on 24th Oct 2011. In order to improve the efficiency, two survey modes were adopted: survey for bright nights and survey for dark nights.  The bright nights is defined as from 8 to 20 nights after new moon, corresponding 13 nights  per lunar month, while the dark nights from 4/5 nights before new moon to 5 nights after full moon, which is 10/11 nights per luna month. Six test nights per month is also arranged from 5 nights to 7 nights and from 20 nights to 22 nights after new moon. In the bright nights, the galactic anti-centre (GAC) area, the galactic disk area and a constant latitude stripe are observed. In the dark nights, the survey of galactic halo part (known as LEGUE, LAMOST Experiment for Galactic Understanding and Exploration, not including GAC?) and extragalactic objects (known as LEGAS, LAMOST ExtraGAlactic Surveys) are carried. The pilot survey will tell us how this telescope can perform and what kind of scientific goal we can achieve during the main survey.

%\section{Observation area for the dark nights}   
\section{Regions of sky for dark night observing}%% Modification 1
\label{sect:Obs}

%The observation for dark nights contains 3 areas: The LEGAS stripes, the low galactic latitude box, and the GD-1 stream area. There are two LEGAS stripes, one is RA between $-315^\circ$ and $60^\circ$ and Dec between $-1.5^\circ$ and $8.5^\circ$, another is RA between $120^\circ$ and $240^\circ$ and Dec between $-0.5^\circ$ to $9.5^\circ$(d?) which were designed by Shen. The low galactic latitude box was selected  using RA between $120^\circ$ and $145^\circ$ and Dec between $6^\circ$ and $14^\circ$, where sagittarius stream and monoceros ring overlap.  The GD-1 area follows the position of GD-1 stream which RA roughly between $134^\circ$ and $220^\circ$ and Dec between $20^\circ$ and $63^\circ$. 

%JLC: "was planned to start" --> "began", added "is planned to"
The LAMOST Pilot Survey began in October 2011 and is planned to continue through the end of Spring 2012; this places some limitations on the accessible range of right ascension (RA). Of course, it is important that targets are available on every night with clear weather. Two ranges in RA that bracket the Galactic plane were selected for the input catalogues: one between $-45^\circ < \alpha < 60^\circ$, and the other between $120^\circ < \alpha < 240^\circ$. To enable studies of Galactic halo stellar populations, the high stellar density regions at low ($b \lesssim 30^\circ$) Galactic latitudes were not included (note that low-latitude stars are included either in the anti-center portion of the survey or the disk portion of the survey, both of which focus on predominantly brighter stars near the disk; see Chen et al. 2012). Due to the telescope site and the unique optical design of LAMOST, the optimal observable sky area is restricted to an area between declinations of $-10^\circ < \delta < 60^\circ$ (Zhao et al. 2012). 

Three regions of sky were selected for dark night observations during the Pilot Survey (see the map in Figure~1). We will denote these the \lq\lq GD-1 area" (shown in blue in Fig.~1), the low Galactic latitude \lq\lq Anticenter Box" (red), and the \lq\lq  LEGAS area" (black). These regions are defined as follows:

\begin{enumerate}

\item The GD-1 area (blue region in Fig.~1) was selected to cover the GD-1 tidal stream (Grillmair \& Dionatos 2006) as much as possible. To do so, a region of $\sim5^\circ$ width was selected surrounding the stream as traced by Willett et al. (2009): 

\begin{equation}
\delta=-864.5161+13.22518\alpha-0.06325544\alpha^2+0.0001009792\alpha^3
\end{equation}

\item The Anticenter Box is defined by $120^\circ < \alpha < 145^\circ$ and $6^\circ < \delta < 20^\circ$. It was selected to overlap some known Milky Way substructures, including the Sagittarius stream, the Anticenter Stream (and/or the \lq\lq Monoceros Ring"; see Li et al. 2012), and the Eastern Banded Structure (Grillmair 2011). 

\item The LEGAS area was selected by the LEGAS group for the study of galaxies and QSOs using LAMOST. Because the surface density of galaxies and QSOs in the observable magnitude range is much smaller than 200 deg$^{-2}$ (the fiber density of LAMOST), we fill the remaining fibers (typically about half of them) with stars selected in the same way as those for the rest of the LEGUE survey. The LEGAS area consists fields near the celestial equator, and is split into two regions (again, to avoid the Galactic plane). One is the South Galactic Cap region (SGC), ranging from $-45^\circ < \alpha < 60^\circ$ and $-1.5^\circ < \delta < 8.5^\circ$, and the other region is in the North Galactic Cap (NGC), ranging from $120^\circ < \alpha < 240^\circ$ and $-0.5^\circ < \delta < 9.5^\circ$. 

\end{enumerate}

Each of these regions was chosen in part to maximize the potential scientific yield of the Pilot Survey data, and in part to provide valuable test data for assessment of LAMOST performance.

The stellar tidal stream of Grillmair \& Dionatos (2006), which has come to be known as \lq\lq GD-1", is one of the nearest known halo substructures. This stream is a narrow feature sweeping across much of the NGC region of the sky. Distances derived from F-type main sequence turnoff stars (at $g$ magnitudes of $\sim18.5$) place GD-1 at 7-10 kpc from the Sun (Willett et al. 2009). With kinematics and distances over a large angular extent, one can fit an orbit to the stream and use this to trace the gravitational potential of the dark halo of the Milky Way (e.g., Koposov et al. 2010, Willett et al, 2009). Another benefit of the GD-1 region for the Pilot Survey is that it spans a large range in declination. As mentioned previously, the telescope performance varies with declination; data from the GD-1 area will provide valuable test data to assess the data quality over a large range of declinations.

\begin{figure}
   \centering
   \includegraphics[width=\textwidth, angle=0]{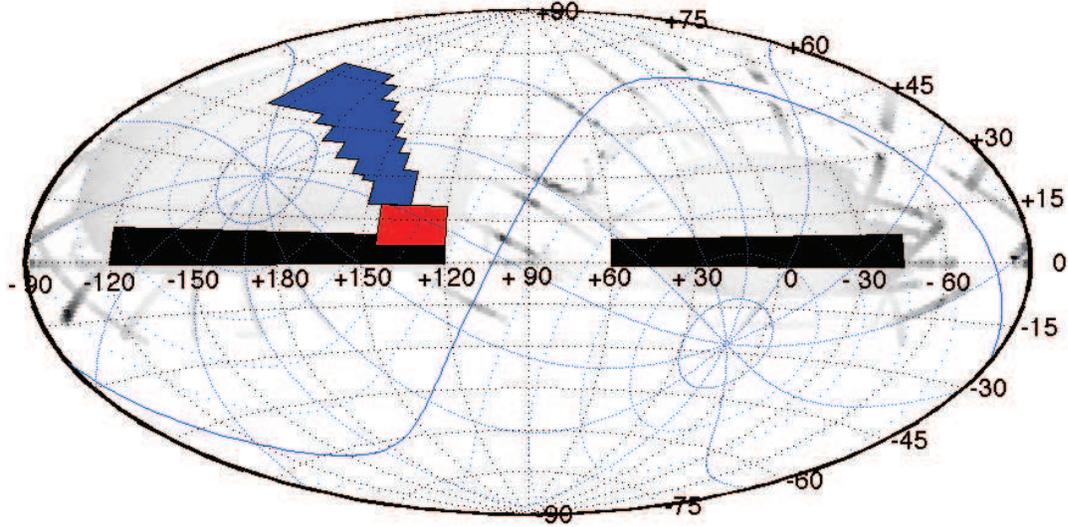}
   \caption{Sky areas targeted on dark nights during the LAMOST Pilot Survey, shown in equatorial coordinates. The figure is centered at (RA, Dec)$_{\rm J2000}$=($90^\circ,0^\circ$). Black dots show the RA, Dec coordinate grid, while blue dots show Galactic longitude and latitude. The thick blue line shows the Galactic plane. The grey background is a stellar density map of stars with $-0.5 < (g-r)_{0} < 0.4$ and $20 < r_{0} <22$ from SDSS DR8 photometry. The LEGAS area (filled black rectangles) has two stripes, one is RA between $-45^\circ$ and $60^\circ$ and Dec between $-1.5^\circ$ and $8.5^\circ$, another is $120^\circ <$ RA $< 240^\circ$, $-0.5^\circ <$ Dec $< 9.5^\circ$.
%    which were designed by Shen (2012, RAA). 
The Anticenter box (filled red box) was selected by $120^\circ <$ RA $< 145^\circ$, $6^\circ <$ Dec $< 20^\circ$, where the Sagittarius stream and ``Monoceros Ring'' substructures overlap. The GD-1 area (filled blue region) follows the position of the GD-1 stream as fit by Willett et al. 2009 (see Equation 1). }
%RA roughly between $134^\circ$ and $220^\circ$ and Dec between $20^\circ$ and $63^\circ$. }
   \label{Fig1}
   \end{figure}

The Anticenter Box (red box in Figure~\ref{Fig1}) was selected for the study of known stellar tidal streams. Obtaining thousands of spectra covering this complicated region of sky is essential for disentangling these numerous features that overlap on the sky. Additionally, these data will probe the Milky Way at moderate latitudes ($20^\circ \lesssim b \lesssim 45^\circ$) near the Galactic anticenter, providing a valuable data set for studies of thin/thick disk structure in addition to the substructures that will be present in the data.

Finally, the LEGAS regions, while selected by the extragalactic group to optimize their science, will provide a valuable data set for Galactic structure studies as well. These two regions (the filled black rectangles in Figure~1) probe a large volume of the high-latitude Milky Way in both the southern and northern Galactic caps. The stellar spectra from these regions can be used to study the overall density structure of the Galactic halo; in particular, these data will provide a means of comparing the stellar density profile in the NGC to that of the SGC. Recent evidence has shown that the Milky Way stellar halo is asymmetric (Newberg et al. 2007; Newberg \& Yanny 2006; Xu et al. 2006, 2007), with excess stars in the north relative to the southern Galactic cap. At least some of this asymmetry is due to the large, cloud-like Virgo substructure (e.g., Newberg et al. 2007; Juric et al. 2008), a $\sim1000$ deg$^{-2}$ stellar overdensity in the northern Galactic hemisphere, centered at $\alpha \sim 180^\circ$ (e.g., Vivas et al. 2006). The LEGAS stripe will yield a large number of spectra of candidate Virgo substructure members. Overall, the LEGAS stripes, while effectively \lq\lq bonus" data for the LEGUE survey, will be an important resource for Galactic structure studies.

% reason why GD-1 area was specially designed is that GD-1 stream is the one of the nearest known halo substructure (7-10 kpc, Willett) and the visible g magnitude of its F-turn off is around 18.5 mag which is also doable for the original technical design of LAMOST.  Apart from testing the performance of the telescope, the pilot survey could also be used to trace the GD-1 stream. With the help of the spectroscopic data, we can better under stand the gravitational potential of the dark halo of the Galaxy. The Anti-center Box area was chosen because this is an area where many substructure overlap (Sagittarius stream, Monocerous ring, and GD-1 stream). We can use the one-year pilot survey to study the structure of this area. The extra area beyond those three regions was also prepared because some plates could be placed near the edge of the area, and we must make sure the whole plate has enough targets to be observed.   
   
%--------------------------------------------------------------------------------
\section{TARGET SELECTION}           

%\subsection{Description of the source}
The fixed-meridian design of the telescope restricts observations to objects within $\sim2$ hours on either side of the meridian plane.  The observational magnitude limit in $r$ band for the Pilot Survey is $r \lesssim 19.5$, which can be achieved by summing 3 exposures of 1800s each. This 1.5-hour exposure time is the standard for the faint plates in the Pilot Survey. 

In order to generate the input catalogue for the dark nights portion of the LAMOST Pilot Survey, we need to find a proper source of high-quality, homogeneous photometry. It must be photometrically complete to the LAMOST magnitude limit ($r\sim19.5$), and must cover much of the northern hemisphere sky accessible from Xinglong Station (the LAMOST site).
% be a whole sky coverage source. 
An ideal source for this is the publicly available Sloan Digital Sky Survey (SDSS) Eighth Data Release (DR8; Aihara et al. 2011), which covers 14,555 square degrees in sky area, mostly in two large contiguous areas near the north and south Galactic caps. SDSS DR8 contains more than 260 million stars, and is 95\% complete to magnitudes of $r\sim22$. For the Galactic halo portion of the LEGUE survey, the five photometric bands ($u, g, r, i,$ and $z$) provided by SDSS offer the possibility of selecting different types of stars of particular interest based on their colors. We thus chose to use SDSS DR8 photometry to create input target catalogs for the relatively high-latitude LEGUE survey.
% for r magnitude, which makes it a perfect choice to be the source of our input catalogue. 
   
\subsection{Color and magnitude target selection criteria}

All stars between $14 < r < 19.5$  were selected from SDSS DR8 as the source of the input catalogue for dark nights. Candidates were selected from the SDSS \lq\lq Star" database (i.e., objects identified as point sources by the SDSS pipelines) using the ``clean" photometry flag to ensure that we obtain only well-measured stars. Unlike brighter magnitude-limited spectroscopic surveys such as RAVE (Steinmetz et al. 2006) or HERMES (Freeman \& Bland-Hawthon 2008), the number of available sources for the faint plates in the LEGUE survey is much larger than the number of objects that can be observed. To deal with this problem, we adopt the Target Selection (TS) algorithm described in Carlin et al. (2012), which implements a way to preferentially select targets of particular types based on any observable quantity (or combinations of observables), while retaining a smoothly-varying, well-understood selection function that samples stars from all regions of parameter space. For LEGUE, we specifically are interested in biasing the target selection to include as many horizontal-branch (and/or blue straggler) stars as possible, as well as over-selection of blue F-type turnoff stars. These categories of intrinsically bright stars of relatively unambiguous luminosity classification are extremely useful for studies of the Galactic halo to large distances. In addition, the selection is biased toward bluish bright stars at brighter magnitudes, from which we hope to identify candidate extremely metal poor stars at bright enough magnitudes for high-resolution spectroscopic follow-up. (Note that these and other LEGUE science goals are outlined in Deng et al. 2012.) We would like to de-emphasize M dwarf disk stars, which would otherwise dominate the sample, but still observe a large sample of M stars. The rough overall targeting goals are the following (note that these actually pertain to the full LAMOST survey; these goals can only be partly accomplished during the Pilot Survey):

%, which would like to achieve the following goals:
\begin{enumerate}
\item Select nearly all stars with $0.1 < (g-r) < 1.0$ and $r < 17$ at high Galactic latitudes, and subsampled at $b < 40^\circ$.
\item Select nearly all stars with $g-r < 0.0$ and $u-g$ colors that suggest they are not quasars.
\item Select a significant fraction of the stars with $0.0 < (g-r) < 1.0$ and $17 < r < 19.5$ and $u-g$ colors that suggest they are not quasars. The bluer side of the color range should be selected with a probability about twice the redder side of the range. The stars should be somewhat evenly distributed in magnitude.
\item Select a large number of M dwarfs at all magnitudes.
\end{enumerate}

As outlined in Carlin et al. (2012), each star from an input catalog can be assigned a probability of being selected for targeting using a function of the form:

\begin{equation}
P_{\rm j, D}= \frac{K_{\rm D}}{\left [ \Psi _{0}\left ( \lambda _{\rm i} \right ) \right ]_{\rm j}^{\alpha}}f_{\rm i}\left ( \lambda _{\rm i} \right )
\end{equation}

\noindent where $\lambda_{\rm i}$ are any observables that are known for all stars in the input catalog (i.e., photometry,
astrometry, or any combination of observed quantities). The $\Psi_0(\lambda_{\rm i})$ term is the statistical distribution function of
the observable $\lambda_{\rm i}$, and $K_{\rm D}$ is a
normalization constant to ensure that the probabilities sum to
one. The function $f_{\rm i} (\lambda_{\rm i})$ can be any smooth function of the observables, and can be used to add emphasis to certain regions of parameter space.
%The "local density" of stars in multidimensional observable
%space is calculated as, for example, the number of stars within some fixed range of  
The \lq\lq local density", $\Psi_0$, was determined for each star in the input catalog
%, we calculated the local density of sources at $(\lambda_1, \lambda_2, ...,
%\lambda_i)$, estimated 
by counting the number of stars $j$ whose
observables satisfy the condition: 

\begin{equation}
\sqrt{\sum_{\rm i} (\lambda_{\rm i}-[\lambda_{\rm i}]_{\rm j})^2} < {\Delta\lambda}
%\sum_{\rm i} \frac{(\lambda_{\rm i}-[\lambda_{\rm i}]_{\rm j})^2}{(\Delta\lambda_{\rm i})^2} < 1
\end{equation}

\noindent where the $\lambda_{\rm i}$ were chosen to be $r$ magnitude and $g-r$, $r-i$ colors. The $\Delta\lambda$ defines the size of the volume in the three dimensional (magnitude, color, color) space over which the density of stars is being counted, and can be thought of as the \lq\lq resolution" of the function $\Psi_0=\Psi_0(r, g-r, r-i)$.
% For example, one could
%determine the density function $\Psi_0=\Psi_0(g,g-r)$ 
We chose $\Delta\lambda = 0.1$, defining the local density as the number of objects found within 0.1 magnitudes of the location of each star in $(r, g-r, r-i)$ parameter space.

%, which would mean using $\Delta g=\Delta (g-r)=0.1$
%mag. This can be extended to any number of the observables to define a
%"density" over multiple parameters; an example would be using
%additional colors, calculating the number of stars within 0.1
%magnitude of ($g, u-g, g-r, r-i$,...).

%, and the candidates are weighted by some power of this "local
%density". 
The exponent $\alpha$ in Equation~2 weights candidates by some power of the local density, and is typically between 0 and 1. When $\alpha=0$, all targets have the same probability of being selected (i.e., it is a random selection), so the selected sample will have the same distribution in the observables used to define the local density as that of the
input sample. To produce a selection that is evenly distributed across the observable space, one can weight the probability for selection by the inverse of the local density, i.e., $\alpha=1$. This type of sample will over-emphasize rare objects in relatively sparsely-populated regions of parameter space.  We examined the actual targets \lq\lq observed" in many simulations of plate selection at different latitudes, and based on the distribution of targets selected we determined that an intermediate case of $\alpha=1/2$ (i.e., weighting by the inverse
square root of the local density) best produced our desired overemphasis of rare objects while retaining many stars from higher-density regions. To accomplish the extra weighting of brighter, $r < 17$ stars desired from point (1) above, we include a linear \lq\lq ramp" function $f(r) = 1 - 1.0*(17.5 - r)$ in $r$ magnitude beginning at r = 17.5 and increasing toward brighter stars with slope of 1.0. Likewise, a ramp in $g-r$ color was applied to overemphasize blue stars as desired in criteria 1-3 above: $f(g-r) = 1 - 2.5*[1.1 - (g-r)]$. Some examples of the effects of these criteria on the color and magnitude distributions of targets will be seen later in this work.

%are over-emphasized. For the LEGUE bright survey targets, we selected
%an intermediate case of $\alpha=1/2$ (i.e., weighting by the inverse
%square root of the local density), which emphasizes the selection of
%rare objects but keeps a large number of stars from higher-density
%regions in the observables. Unlike the dark nights survey of faint
%targets, we did not add any linear weights in color or magnitude to
%the bright survey selection criteria.

%Through many tests, a combination of parameters was determined:
%
%\begin{enumerate}
%\item $\alpha$ = 1/2, .i.e. weighting by the inverse square root of the local density
%\iteXXXm Linear ramp in g - r, beginning at g - r = 1.1 and increasing blueward with a slope of 2.5.
%\item  Linear ramp in r magnitude, beginning at r = 17.5, increasing toward brighter stars with slope of 1.0.
%\end{enumerate}

\section{PLATE DESIGN}

An important consideration in LAMOST survey design is the requirement that each plate must have a bright star at its center, which is fed to the Shack-Hartmann (SH) wavefront sensor to derive active optics corrections. Without the active corrections, the images of stars will be wildly out of focus in the focal plane.
%Besides input catalogue, the selection of Shack-Hartmann stars (SH) is very critical. 
SH stars must be brighter than 8th magnitude in $V$-band in order to provide sufficient flux for high-frequency active mirror distortions.
One goal of the Pilot Survey is to cover as much sky area as possible, with each plate observed only one time. This theoretically requires the Shack-Hartmann stars to be separated by roughly $5^\circ$ (the diameter of a LAMOST plate) from each other. The distribution of all potential Shack-Hartmann stars in the dark nights survey region from the Hipparcos (Perryman et al. 1997) catalog is shown in Figure~2 as yellow dots, with red circles of $5^\circ$ diameter around each SH star representing possible LAMOST plates. Most of the desired regions on the sky can be easily covered with available SH stars, but there are a few small high-latitude areas that will be inaccessible to the survey. Based on the calculation of the site conditions and the available dark time during the period of the Pilot Survey (see, e.g., Deng et al. 2012, Yao et al. 2012), around 30 dark night plates will be observed; Approximately 15 of these will be in the Anticenter Box (the red area in Figure~1), and 15 in the GD1 area. Figure~3 shows a sample tiling of 15 plates in each of these regions, centered on known $V<8$ SH stars. The additional yellow asterisk symbols are unused SH-star candidates; these show that contiguous coverage of these sky areas can be easily achieved in the main LAMOST survey. How the plates for dark nights that designed for LEGUE area will be described below. 

\begin{figure}
   \centering
     \includegraphics[width=\textwidth, angle=0]{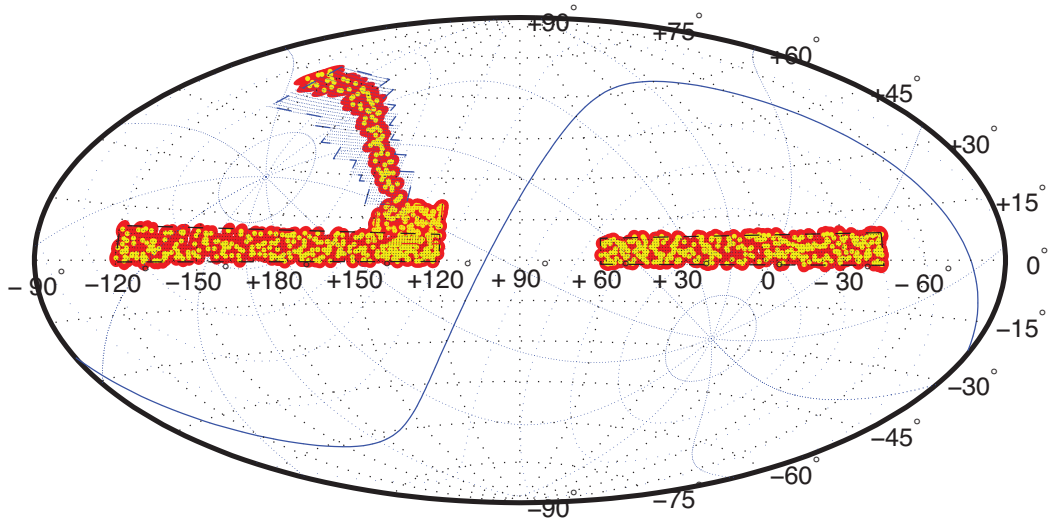}
   \caption{Sky coverage for all possible plates for dark nights: As shown in Figure1, blue is GD1 area, red is Anti-center Box, and two black rectangles are SGC and NGC areas, respectively. The red circles show all the possible plates while the Shack-Hartmann star is selected to be the center for each plate, seen as yellow dots.}
   \label{Fig2}
   \end{figure}

\begin{figure}
   \centering
   \includegraphics[width=.8\textwidth, angle=0]{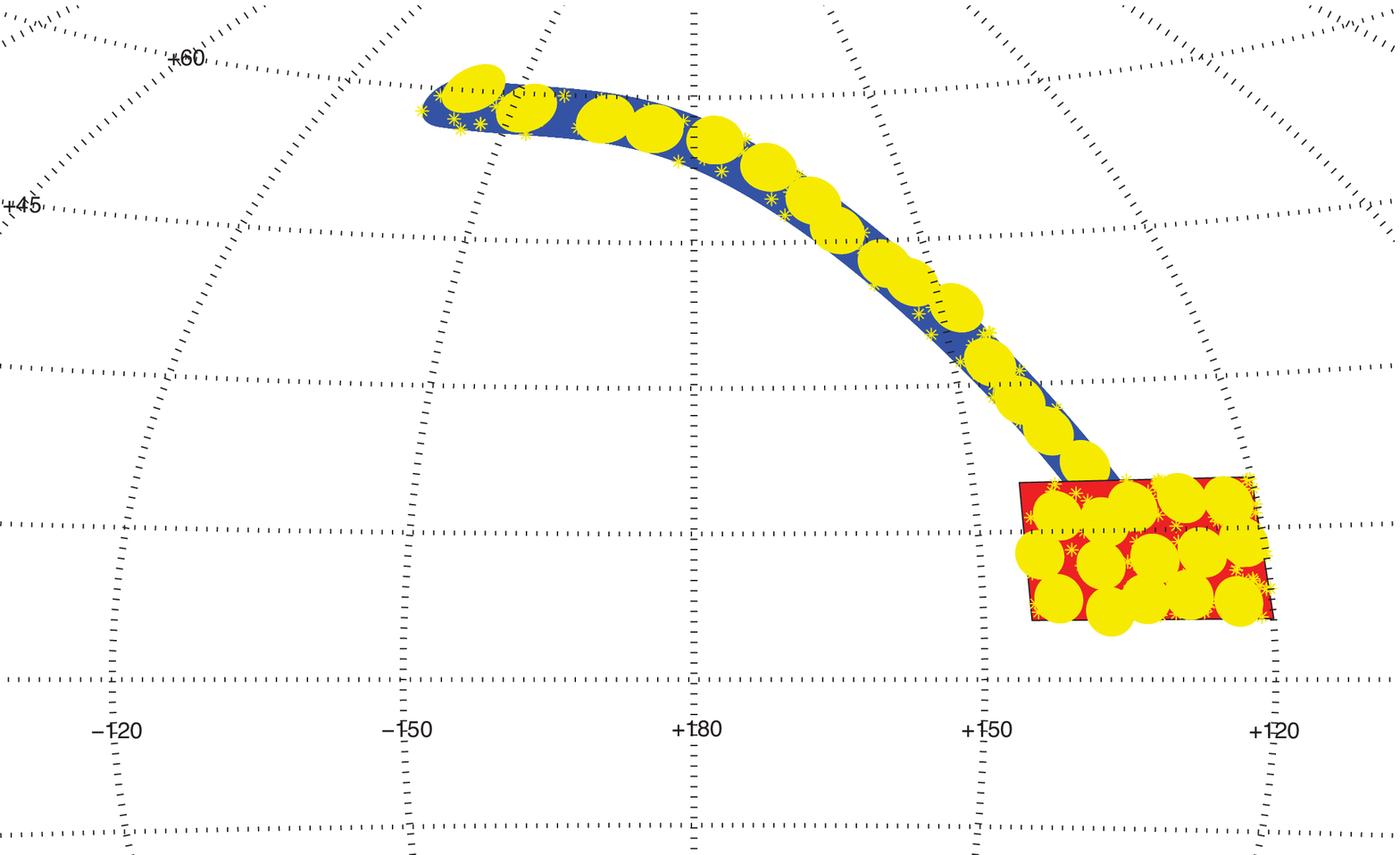}

   \caption{Plates designed for GD1 area and Anti-center Box in equatorial coordinates. The red area is the anti-center red box while the blue area is the GD-1 area. The yellow circles are designed plates to cover as much as sky area with limited plate number. It could be different in the real observation due to other factors such as weather condition, observing time window and so on. The yellow asterisks are unused SH-star candidates; these show that contiguous coverage of these sky areas can be easily achieved in the main LAMOST survey with a SH limiting magnitude of $V=8$. }
   \label{Fig3}
   \end{figure}

For the Pilot Survey, we extracted all photometry from SDSS DR8 in each of the four regions (Anticenter Box, GD1 Area, LEGAS SGC, and LEGAS NGC; i.e., the filled regions in Figure~1) to be observed during dark nights. We then applied our general target selection algorithm to these large catalogs, assigning selection probabilities to stars using the TS algorithm from Carlin et al. (2012), with the parameters given in Section~3.1, until a density of 600 stars deg$^{-2}$ was reached. This input catalog density was used to satisfy the requirement of the LAMOST fiber assignment software of a catalog containing three times the number of stars to be assigned; LAMOST has 200 fibers deg$^{-2}$ in the focal plane, so 600 stars deg$^{-2}$ is required for fiber assignment. This means that each plate selected in these areas will consist of an input catalog of $\sim12000$ stars, from which $\sim4000$ will be selected for observation.

One complication that arises is that during the pilot survey, the LAMOST fiber assignment program (Survey Strategy System; SSS) is not able to deal directly with the selection probabilities as calculated. Instead, SSS uses an integer priority value (currently constrained to values 0-99), with lower numbers equaling higher priority for assignment. When assigning fibers, SSS considers all stars available to each fiber, and if possible assigns the one with the highest priority among those available. If all stars in an input catalog have equal priority, then the resulting plate will have targets with a roughly uniform spatial distribution, because SSS will simply assign the target nearest the ``home" position for each fiber. Thus to ensure that our assignment probabilities have the desired effect, we must assign each star in the input catalog a priority flag according to the process outlined in Section~2 of Carlin et al. (2012). For the dark nights survey we chose to assign priorities 1-80, reserving the remaining values in case they were needed for other purposes. For each square degree of sky, the 600 targets deg$^{-2}$ were divided into 80 bins with 600/80 stars in each. We loop over the priority bins, selecting stars using the selection probabilities until each priority bin has the desired number of stars. In this way, the high probability stars are most likely to have high priority for assignment (i.e., low integer priority value) because they have a higher likelihood of being selected near the beginning of the process.  

The large input catalogs with 600 stars deg$^{-2}$ target density were given to the LAMOST observing specialists together with the list of possible Shack-Hartmann stars in these regions. The observing group can then decide for each night which SH stars the plates will be centered on, and use SSS to allocate stars to fibers for observation. Operating in this mode is preferable to designing each plate in advance because it provides the observers some flexibility in scheduling observations. Because LAMOST is limited to observing near the meridian, only limited regions (in right ascension) are available at any given time, making contingency plans essential for efficient survey operations. Also, the effects of atmospheric refraction can slightly change the objects available to each fiber.

%Applying the TS algorithm and using DR8 photometry as the source catalog, we generated four sets of big input catalogue for 4 observing area, each containing 600 stars/square degree. This means each plate in such area will cover over 12000 stars, which is 3 times as the number of stars will be observed. We then gave the input catalogue to the observing group together with the list of SH stars . The observing group made the observation plan with many backup plates so as to fit the real time observation. It is because the time used to adjust the mirror angle and fibre position could vary in some degree, and they must make sure there will always be plates to be observed. They then decided which SH star the plate would centre on and which stars would be observed using Survey Strategy System (SSS), a software that rapidly arranges observing plans in the most efficient way.

Figures 4-7 show examples of the color and magnitude distributions of the input data and selected stars in two plates at high and low Galactic latitudes. At high Galactic latitude, we present a sample plate designated F5593003, centered at $(l, b) = (132.6^\circ, 60.2^\circ)$. In Figure 4 we show a comparison of the $r$-magnitude and $g-r$ distributions (left and middle panels, respectively) between the source (DR8) photometry (blue histograms), the input catalog (i.e., candidates given to the fiber assignment algorithm; red lines), and the stars selected for observation (in black). The histograms have been normalized by the total number of stars in each set to produce something equivalent to a probability distribution (i.e., the probability of any randomly selected star falling within one of the bins is equal to the bin height). The right panel shows the fraction of available DR8 (source) stars in each color bin that were selected for the input catalog. This clearly illustrates the overemphasis of blue objects -- nearly all of the stars bluer than $(g-r) = 0.0$ are selected by the target selection algorithm for input to the SSS fiber assignment code. Hess diagrams for the same three datasets (source, input, and observed) in F5593003 are seen in Figure~5. From these figures, one can see that the selected targets contain more bright, blue stars than the input catalogs, having de-emphasized the faint, red (predominantly M-dwarf) stars. However, because the stellar density at high ($b \sim 60^\circ$) latitudes is rather low, a large fraction of available stars are actually assigned to fibers in F5593003. This is not the case at lower latitudes -- as an example, we show the plate F5591504, centered at $(l, b) = (207.0^\circ, 23.3^\circ)$, in Figures 6 and 7. Note that in this low-latitude field with much higher stellar density, the difference between the overall input distributions and the observed targets is more dramatic. This is simply a reflection of our target selection scheme and our desired overemphasis on bright, blue targets.

%For high Galactic latitude area, we use F5593003 centered at (l,b)=(132.60,60.19) as an example to show the comparison between the source, the input catalogue and the final observation (Figure 4). The corresponding Hess diagrams are shown in Figure 5.  We can see through TS and SSS, the statistic profile of r magnitude and g-r will both vary slightly because low stellar density at the high latitude area which can cause the offset from statistic profile. SSS itself also sets an 5 magnitude limit between the faintest and the brightest objects. That is the reason why stars with r magnitude brighter the 14.5 or fainter than19 were cut out, which is also a problem we must solve in the coming main survey. One would expect to have stars between 14.5 and 19.5 in r magnitude other than between 14 and 19, for the number of stars with $r\sim14-14.5$ mag is much less than $r\sim19-19.5$  mag. While for low Galactic latitude, we use F5591504 as a contrast (Figure 6 \&\ 7). The profiles of both r magnitude and g-r are quite similar between the stars selected by TS and those selected by SSS, which is also expected for the Pilot Survey. We then can use the distribution of observed stars to inverse the original distribution of the stars in the sky area, which is also critical for a whole sky survey.   

\begin{figure}
   \centering
   \includegraphics[width=\textwidth, angle=0]{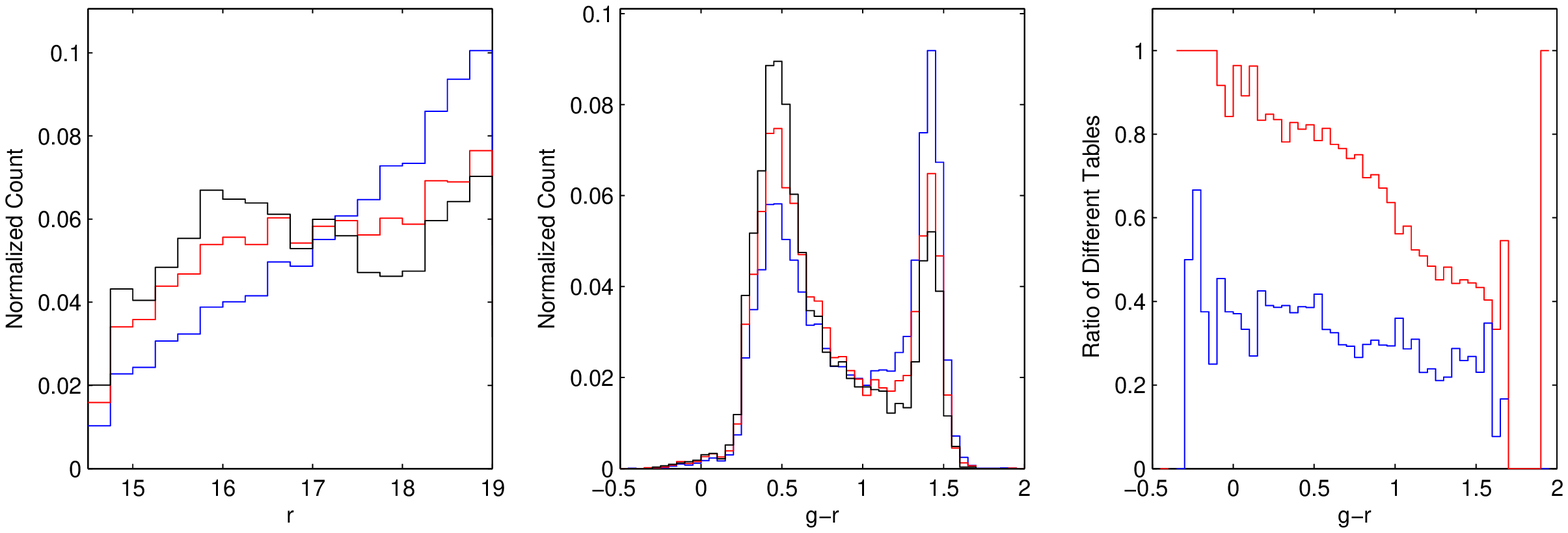}
   
   \caption{Distribution in $r$ magnitude (left panel), $g-r$ (middle panel) and proportion of input catalog to source (DR8) catalog for stars selected in the field of F5593003 centered at (RA, Dec)$_{\rm J2000}$=($184.21^\circ,56.28^\circ$). For the left and middle figures, the solid line shows the distribution for all stars (20406 stars) in the field of view, the dash dot histogram shows the 12512 stars selected using the TS algorithm and the dotted line is the histogram of 3744 stars selected by SSS and will be put into real observation. In the right panel, the solid line shows the proportion of input catalog to source (DR8) catalog versus color, while the dash dot line, SSS selected catalogue to input catalogue.}
   \label{Fig4}
   \end{figure}

\begin{figure}
   \centering
      \includegraphics[width=\textwidth, angle=0]{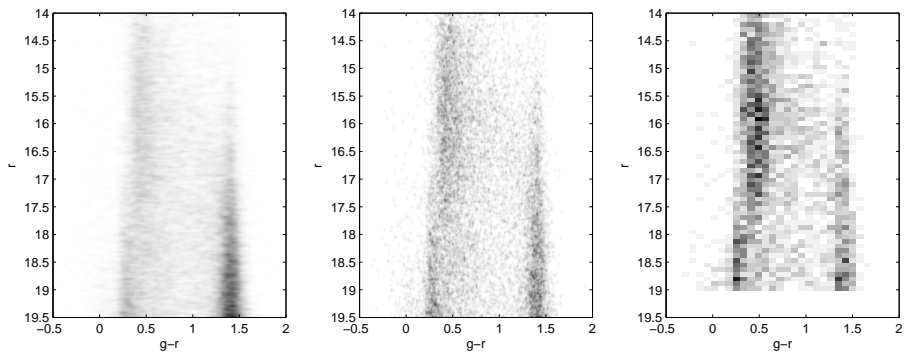}

   \caption{Color-magnitude Hess diagram of stars in the plate F5593003. Left panel: All stars in the field of view. Middle panel: Stars selected applying TS algorithm. Right panel: Stars selected by SSS.}
   \label{Fig5}
   \end{figure}

%\subsection{From input catalogue to real observation}
%\label{sect:observe}

%We give the input catalogue to the observing group together with the list of SH stars . The observing group will make the observation plan in advance which contains many backup plates so as to fit the real time observation. It is because the time used to adjust the mirror angle and fibre position could vary in some degree, and they must make sure there will always be plates to be observed. They then decide which SH star the plate will centre on and which stars will be observed using Survey Strategy System (SSS), a software that rapidly arranges observing plans in the most efficient way. Again, F5593003 is used to compare the input catalogue and the final observation (Shown in Figure 6). The Hess diagram of the stars in real observation is shown in Figure 7.  We can see through SSS, the statistic profile of r magnitude will vary slightly because the SSS set an magnitude limit between the faintest and the brightest objects. That why stars with r magnitude fainter than19 were cut out. This is a problem we must solve in the coming main survey. One would expect to have stars between 14.5 and 19.5 in r magnitude other than between 14 and 19, for the number of stars with $r\sim14-14.5$ mag is much less than $r\sim19-19.5$  mag. The distribution of g-r and the Hess diagram remain very similar. 

\begin{figure}
   \centering
   \includegraphics[width=\textwidth, angle=0]{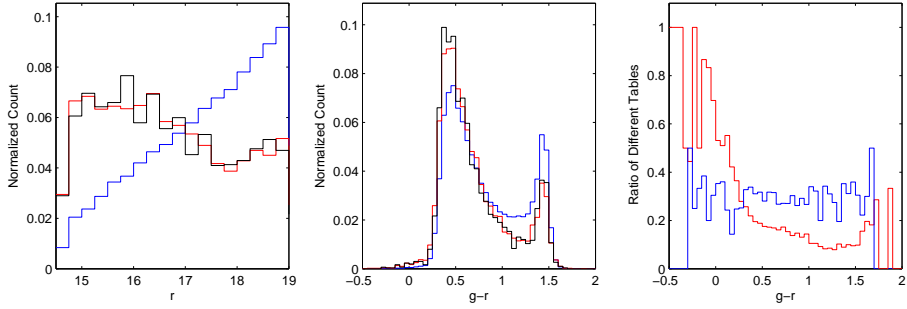}

  \caption{Distribution in $r$ magnitude (left panel) and $g-r$ (middle panel) for stars selected in the field of F5591504 centered at (RA, Dec)$_{\rm J2000}$=($121.54^\circ, 15.35^\circ$). The lines are defined as in Figure 4 (78\,629 stars in the field of view, 11\,758 selected from TS and 3902 from SSS). In the right panel, same as fig 4, the solid line shows the proportion of input catalog to source (DR8) catalog versus color while the dash dot line shows the proportion of SSS selected catalogue to input catalogue.}
   \label{Fig6}
   \end{figure}

\begin{figure}
   \centering
      \includegraphics[width=\textwidth, angle=0]{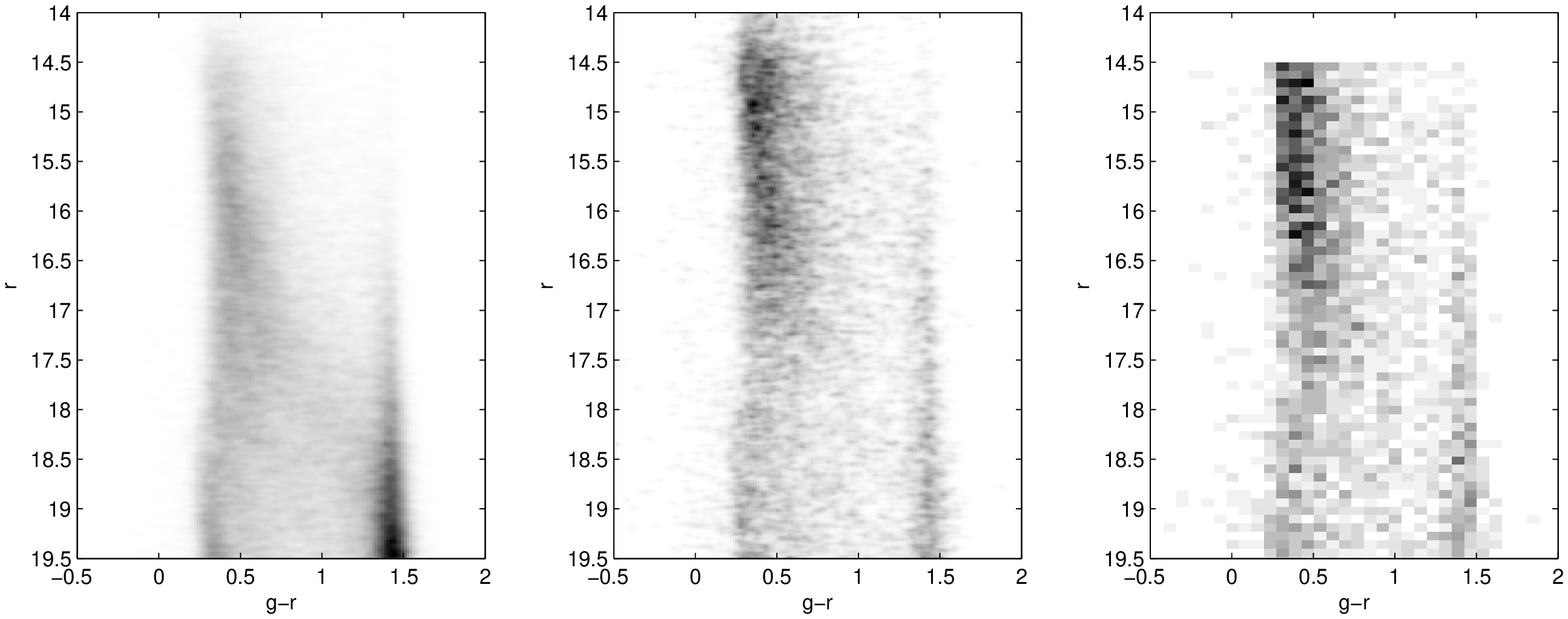}

   \caption{Hess diagram of F5591504 selected using SSS. Left panel: All stars in the field of view. Middle panel: Stars selected applying TS algorithm. Right panel: Stars selected by SSS.}
   \label{Fig7}
   \end{figure}

\section{Summary}
\label{sect:summary}

We have presented details about the design of the LEGUE dark nights portion of the recently completed LAMOST Pilot Survey. The survey of faint stellar targets designed for clear, dark nights consists of three regions of sky. The \lq\lq Anticenter Box" region spanning $120^\circ <$ RA $< 145^\circ$ and $6^\circ <$ Dec $< 20^\circ$, was chosen because a number of known Galactic halo substructures are present in this region. The \lq\lq GD-1 Area" is a region of $\sim5^\circ$ width centered on the narrow tidal stream traced across the northern SDSS footprint by Grillmair \& Dionatos (2006). Finally, the LEGAS areas consist of two strips near the celestial equator that share targets with the extragalactic survey. These regions were selected to maximize the science impact of the Pilot Survey while providing data on survey performance, including a range of declinations and Galactic latitudes.

The targets for the dark nights survey were selected from SDSS DR8 photometry using the algorithm outlined by Carlin et al. (2012). In particular, objects in relatively sparsely-populated regions of ($r, g-r, r-i$) phase space were overemphasized, and additional emphasis was placed on objects toward blue colors and bright magnitudes. Each of these choices is motivated by scientific goals of the LEGUE collaboration. The effects of our target selection process were illustrated for a low and high latitude field by comparing the magnitude and color distributions of stars in the input (SDSS) catalog to those selected for observation. 

In total, the dark nights portion of the LAMOST Pilot Survey should yield $\sim10^6$ stellar spectra. These data will provide a valuable science resource as well as serving as test data for refinement of the targeting and fiber assignment process in the main LAMOST/LEGUE survey.

\begin{acknowledgements}
We thank the referee, Joss Bland-Hawthorn, for helpful comments on the manuscript. This work is partially supported by National Natural Science Foundation of China (NSFC) through grant No.10573022, 10973015 and 11061120454, and the US National Science Foundation (US NSF) through AST grant 09-37523. The Chinese Academy of Sciences (CAS) is acknowledged for providing initial support through grant GJHZ200812.
\end{acknowledgements}

\label{lastpage}

\end{document}